\def\year{2021}\relax
\documentclass[letterpaper]{article} 
\usepackage{aaai21}  
\usepackage{times}  
\usepackage{helvet} 
\usepackage{courier}  
\usepackage[hyphens]{url}  
\usepackage{graphicx} 
\usepackage{amsmath,amssymb,amsfonts}
\usepackage{algorithmic}
\usepackage{booktabs, multicol, multirow}
\usepackage{textcomp}
\usepackage{natbib}  
\usepackage{caption} 
\title{Towards Full-line Code Completion with Neural Language Models}
\author{
    Wenhan Wang,\textsuperscript{\rm 1}
    Sijie Shen, \textsuperscript{\rm 1}
    Ge Li, \textsuperscript{\rm 1}
    Zhi Jin \textsuperscript{\rm 1} \\
}

\affiliations{
    \textsuperscript{\rm 1} Key Laboratory of High Confidence Software Technologies (Peking University), MoE; Software Institute, Peking University, 100871, P. R. China \\
    wwhjacob@pku.edu.cn, sjshen@pku.edu.cn, lige@pku.edu.cn, zhijin@pku.edu.cn
}

\begin{document}

\maketitle

\begin{abstract}
A code completion system suggests future code elements to developers given a partially-complete code snippet. Code completion is one of the most useful features in Integrated Development Environments (IDEs). Currently, most code completion techniques predict a single token at a time. In this paper, we take a further step and discuss the probability of directly completing a whole line of code instead of a single token. We believe suggesting longer code sequences can further improve the efficiency of developers. Recently neural language models have been adopted as a preferred approach for code completion, and we believe these models can still be applied to full-line code completion with a few improvements. We conduct our experiments on two real-world python corpora and evaluate existing neural models based on source code tokens or syntactical actions. The results show that neural language models can achieve acceptable results on our tasks, with significant room for improvements.
\end{abstract}

\section{Introduction}
Code completion has become an essential feature of Integrated Development Environments (IDEs). It speeds up the process of software development by suggesting the next probable token based on existing code. The main goal of most existing code completion systems is to suggest accurate variables, arguments, or APIs to developers. Recently, along with the development of deep learning technologies and easy-to-acquire open-source codebases, researchers have started to tackle code completion by learning from large-scale code corpora.

In this paper, we define a new code completion task: full-line code completion. Given a partially completed code snippet, full-line code completion requires predicting the next line of code, different from traditional code completion which only predicts the next code element. Figure 1 shows a motivating example for our task. To complete the last line in Figure 1, traditional code completion needs to predict at least six times separately, and each time the developer needs to choose the correct token. But if we generate the entire line simultaneously, even if the prediction is only partially correct, the developer can correct the code line with fewer operations. 

Currently, the most popular technique in the research area of code completion is language models, especially neural language models. Neural language model is a powerful tool for predicting the next token given a token sequence, which naturally fits the scenario of code completion. Recent researches have shown that large-scale neural language models like GPT-2 \cite{radford2019language} are capable of generating long text, which brings the potential of code sequence generation. 

One of the key challenges in full-line code generation is to guarantee the syntactical correctness of the generated code.  To tackle this challenge, we draw lessons from past researches on semantic parsing. We adopted a widely used framework for syntax-based code generation, which converts the generation of a code snippet into generating its abstract syntax tree (AST) with a sequence of construction actions.

We conduct experiments on two public Python datasets that contain Python files crawled from Github repositories. One dataset is in Python2, and the other one is in Python3. We evaluate the performance of the state-of-the-art approach for traditional code completion, along with a group of neural language models. Our results show that on both datasets, Transformer language models outperform RNN-based models, which is consistent with past researches in language modeling. We also find that syntax-based approaches do not outperform token-based approaches, indicating that directly applying techniques in syntax-based code generation into full-line code completion can be ineffective.

\begin{figure*}[h] 
\centering 
\includegraphics[height=5cm, width=17cm]{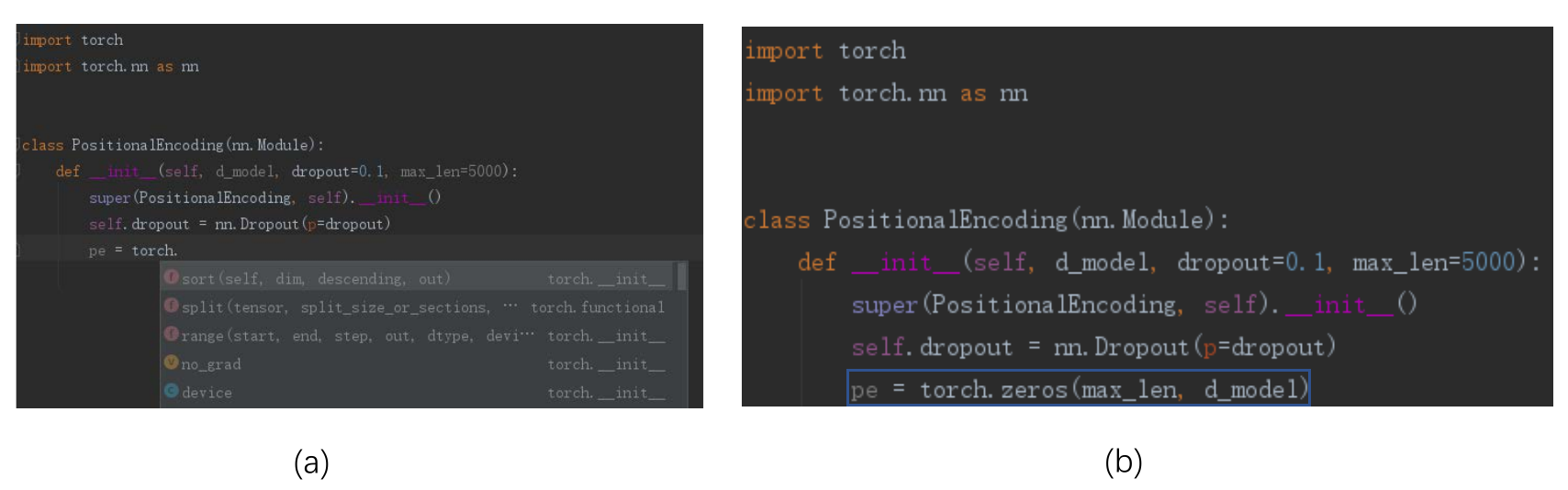} 
\caption{An example showing the difference between traditional code completion (a) and full-line code completion (b)}
\end{figure*}

The main contributions of this paper are summarized as follows:

1) We propose a novel code completion task: full-line code completion and build datasets for this task.

2) We evaluate state-of-the-art models used in traditional code completion and a group of neural language models on our datasets.

3) We analyze the performance of plain token sequence-based language models versus syntax-based language models, and discussed the effectiveness of incorporating syntax information in full-line code completion and some possible improvements in the future.

\section{Related Work}
\subsection{Code Completion with Language Models}
Since \cite{hindle2012naturalness} have found out that source code shares mutual statistical properties with natural language, researchers started using statistical language models on source code. Most early researches use the N-gram language models to model source code. \cite{tu2014localness} pointed out that source code has a unique property of "localness," which could not be captured by the traditional N-gram model. \cite{hellendoorn2017deep} introduced an improved cache-based N-gram model to address the localness and unlimited vocabulary in source code. Their evaluation results on code completion showed that their model outperformed existing statistical and neural language models. \cite{raychev2016probabilistic} proposed a probabilistic language model based on decision trees and domain-specific grammars. They performed experiments on predicting AST nodes rather than directly performing on source code.

Recently, deep learning-based language models have been applied to source code for the code completion task. \cite{liu2016neural} proposed a code completion model based on an LSTM network. \cite{bhoopchand2016learning} proposed an RNN model with a pointer mechanism aiming at copying tokens from the past. Similarly, \cite{li2018code} proposed a pointer mixture network to address the Out-of-Vocabulary (OoV) issue. In predicting the type of the next AST node, their model outperforms \cite{raychev2016probabilistic} on both Python and JavaScript datasets. \cite{karampatsis2020big} use a GRU language model for code completion and use byte-pair encoding (BPE) \cite{sennrich2016neural} to overcome the open vocabulary problem. In the above works, RNN language models are adopted to model the programs. However, RNNs share a common weakness: they cannot efficiently capture the long-term dependencies in sequential data \cite{khandelwal2018sharp}. An efficient way of mitigating long-term dependency problems in neural network language models is to use the Transformer \cite{vaswani2017attention} model. \cite{radford2018improving} first use Transformer to build autoregressive language models and achieved improvement over LSTM-based models on various natural language classification and generation tasks. For code completion, \cite{liu2019self} adopted Transformer-XL \cite{dai2019transformer} on the AST node completion task, and achieved state-of-the-art results.

\subsection{Syntax-based code generation}
The task of automatically generating program code with deep neural networks has been discussed in the natural language processing community as a part of semantic parsing researches.  One of the key challenges of code generation is to generate syntactically valid code snippets. \cite{dong2016language} proposed SEQ2TREE, which uses a tree-structured LSTM to directly generate ASTs, but this model cannot guarantee the validness of generated trees. \cite{yin2017syntactic} addresses this issue by converting the generation of a code snippet into applying a sequence of actions defines by the grammar of the programming language. An action either apply a grammar production rule or emit a terminal token. \cite{rabinovich2017abstract} first introduced abstract syntax description language (ASDL) \cite{wang1997zephyr} grammar into code generation by using a tree-structured decoder. \cite{yin2018tranx} proposed tranX, which further improved the code generation framework in \cite{yin2017syntactic} by replacing the context-free grammar used to generate action sequences into ASDL grammars and expand the framework into a wider range of languages including regular expressions, Python3 and SQL.

\section{Task Definition}

In this section we will describe the task of full-line code completion in details and discuss its difference with traditional code completion.

Given a sequence $S=w_{1},w_{2},...,w_{n}$, language models measures the probability of $S$ by:
\begin{equation}
    P(S)=\prod_{t=1}^{n}P(w_{t}\mid w_{1}w_{2}...w_{t-1})
\end{equation}
Here $P(w_{t}\mid w_{1}w_{2}...w_{t-1})$ can be modeled with a statistical language model or neural network language model. This probability naturally fits the task of code completion, which is predicting the next token in a code snippet.

In full-line code completion, instead of predicting the next single token, the model predicts a sequence of tokens that form a complete statement. Given a partially complete code context $c_{1}c_{2}...c_{k}$, we need to generate a statement composed of a sequence of tokens $s_{1}s_{2}...s_{n}$. If we use a language model to perform full-line code completion, the model need to predict
\begin{equation}
    P(s_{1}s_{2}...s_{n}\mid c_{1}c_{2}...c_{k})=\prod_{t=1}^{n}P(s_{t}\mid c_{1}c_{2}...c_{k},s_{1}s_{2}...s_{t-1})
\end{equation}

Next, we will specify the granularity of "full line" in our task. Roughly, we can take a single line of code as a statement. But this brings an issue, which is in Python we can use a line continuation symbol to make a statement to cover several lines. To solve this issue, we use the Python official library {\fontfamily{\ttdefault}\selectfont tokenize} \footnote{https://docs.python.org/3/library/tokenize.html} to split programs into lines. Generally, there exist two types of code lines in full-line code completion:
\begin{itemize}
    \item Simple statements which implements a intact action, e.g. assignment, return, assertion...
    \item The declaration of a compound statement. e.g. declarations for functions, loops, {\fontfamily{\ttdefault}\selectfont If} statements...
\end{itemize}

\section{Our Approach}
In this section, we will describe the models for full-line code completion in detail. First, we introduce a framework for generating code lines with neural language models. Then we describe the approach of generating lines of code following the syntax in ASTs. 

\subsection{Neural Model for Code Completion}
In this paper we perform neural language models in GRU and Transformer in our experiment. Our Transformer language model follows the architecture of GPT and GPT-2 \cite{radford2018improving,radford2019language}:

\begin{flalign}
    h_{0}&=W_{e}\cdot C+W_{p} &\\
    h_{l}&=transformer\_block(h_{l-1}), \forall l\in [1,N] &\\
    P(u)&=softmax(h_{N}\cdot W_{e}^{T}) &
\end{flalign}

Where $C$ is the code context, $W_{e}$ is the token embedding matrix and $W_{p}$ is the position embedding matrix. The hyperparameters of our model are listed in Table 1. We train the model like traditional language models and maximize the log-likelihood of:
\begin{equation}
    L=\sum_{t}^{}\textrm{log}P(a_{t}\mid a_{1},a_{2},...,a_{t-1})
\end{equation}

Where $a_{t}$ is the token at timestep $t$ in the input program. We do not follow the training procedure in GPT-like models \cite{radford2018improving,radford2019language} which cut input sequences into equal length. Instead, we perform traditional batched training with padding. We assume that splitting action sequences will destroy the syntactical dependencies within a program file, so we keep the input sequences intact in our experiment. Accordingly, as the length of input sequences greatly varies, instead of applying learned position embeddings, we apply the sinusoidal position embedding in the original Transformer \cite{vaswani2017attention}. During inference, we apply beam search to keep candidate sequences with the highest probabilities.

To prepare the action sequence for our full-line code completion task, we need to mark the end of each line in order to terminate the generation after a full line is completed. To achieve this, we manually add an end of line token '$<$eol$>$' after each line in a source code file.

\begin{table}[htbp]
\caption{The hyperparameters of our SG-GPT model.}
\begin{center}
    \begin{tabular}{lll}
    \toprule
    \multicolumn{1}{l}{Hyperparameter} & \multicolumn{1}{l}{Value} \\
    \midrule
    Dimension of self-attention layer $d_{model}$ & 128\\
    Dimension of embeddings $d_{embed}$ & 128\\
    Dimension of the feedforward layer $d_{ff}$ & 512\\
    Number of attention heads $N_{head}$  & 4\\
    Number of Transformer layers $N_{layers}$  & 4\\
    Maximum length of input  & 1500\\
    Dropout keeping probability & 0.9\\
    \bottomrule
    \end{tabular}%
\label{tab1}
\end{center}
\end{table}

Among all types of code tokens, we take an additional step to handle tokens with string type. The contents of strings vary drastically among different programs and often have little relevance to its context, so it is nearly impossible for a language model to suggest an accurate string. Besides, suggesting a wrong string to programmers usually have a negative effect on the user experience. So we mask out all the strings in our data during experiments by replacing all strings in our datasets with a unified token '$<$str$>$'.

\subsection{Syntax-based Code Completion}
In order to make sure the generated code line is syntactically correct, a promising approach is converting the generation procedure of a line into generating its partial AST. In this paper, we adopt the parser of TranX \cite{yin2018tranx} which can decompose an AST into a sequence of tree-constructing actions following an abstract syntax description language (ASDL) \cite{wang1997zephyr} grammar.

\begin{figure*}[h] 
\centering 
\includegraphics[height=3cm, width=15cm]{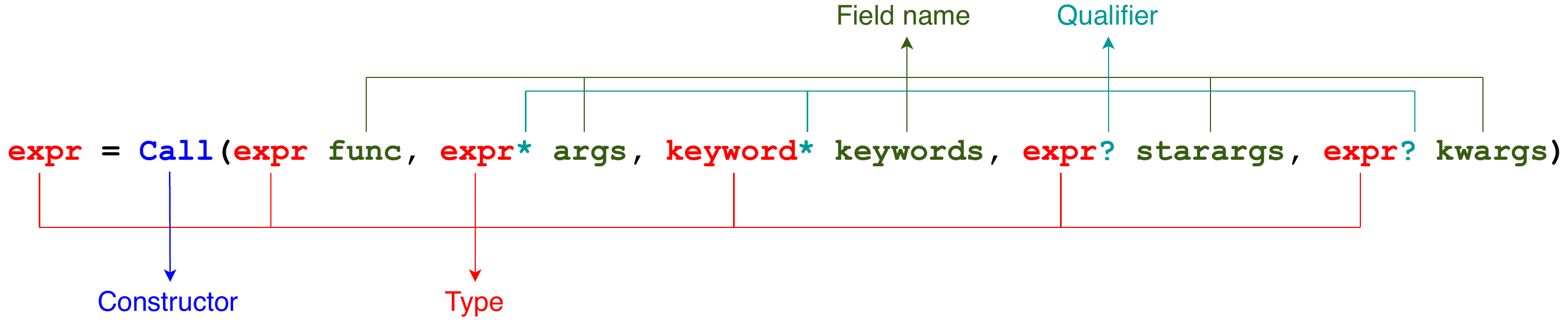} 
\caption{An example of a production rule from the ASDL grammar of Python2.}
\end{figure*}

An ASDL grammar can be seen as an extension of traditional context-free grammar (CFG). Figure 2 shows a typical example of a production rule in the ASDL grammar of Python. Similar to CFG, the values on the right side of an ASDL production are the children of the value on the left. An ASDL grammar has two basic constructs: types and constructors. For example, in Figure 2 a value of type {\fontfamily{\ttdefault}\selectfont expr} can be constructed by a {\fontfamily{\ttdefault}\selectfont Call} constructor. Some constructors have a sequence of fields that describe the type of values associated with a constructor. In this example, the {\fontfamily{\ttdefault}\selectfont Call} constructor denotes function call expressions, and has five fields: func, args, keywords, starargs and kwargs. Each field in the constructor also has a type that specifies the type of value the field can hold. Apart from the type information, each field also has a qualifier (single, optional ? and sequential $*$), which specifies the number of children the field can hold. For example, in Figure 2 The {\fontfamily{\ttdefault}\selectfont func} field can only contain one value, and field {\fontfamily{\ttdefault}\selectfont args} can contain an arbitrary number of values.

An AST is generated in top-down, left-to-right order. The action sequence is composed of three types of actions:
\begin{itemize}
    \item \textbf{APPLYRULE}: apply a production rule from the programming language grammar to expand a non-terminal AST node.

    \item \textbf{GENTOKEN}: generate an exact value for a terminal node, i.e., variable, constant, API call, etc.

    \item \textbf{REDUCE}: a \textbf{REDUCE} action marks the completion of the generation of children for a field with optional (?) or multiple ($*$) qualifier.
\end{itemize}

With these three types of actions, a Python code snippet can be converted into an action sequence and back into source code unambiguously. Figure 3 shows an example of converting a Python statement to an action sequence. We can train and evaluate neural language models on action sequences similar to source code token sequences.

During each timestep of generating the next code line, tranX will inspect the already generated actions and select a set of valid actions at the current timestep. Before applying the softmax function to predict the next action, we apply a mask on the hidden state of the language model to set the positions of all invalid actions to 0. This ensures that the generated action sequence can always be converted back to source code.

\begin{figure*}
  \begin{minipage}[h]{0.45\linewidth}
    \centering
    \includegraphics[width = \linewidth]{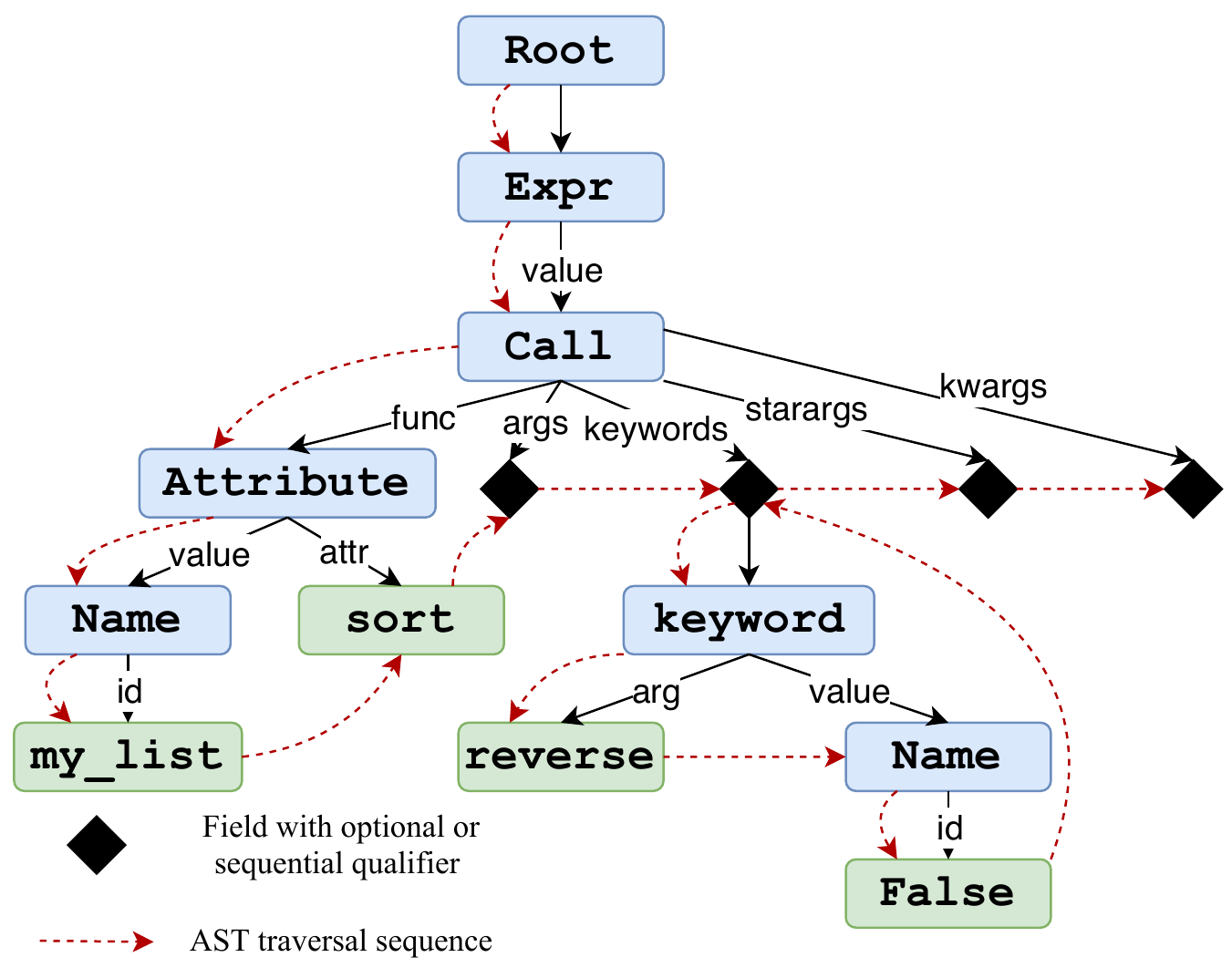}
  \end{minipage}\quad
  \begin{minipage}[h]{0.45\linewidth}
    \centering
    \begin{tabular}{ll}
        \toprule
        \multicolumn{1}{l}{Timestep} & \multicolumn{1}{l}{Action} \\
        \midrule
        1 & \textbf{ApplyRule}[stmt -\textgreater Expr(expr value)]\\
        2 & \textbf{ApplyRule}[expr -\textgreater Call(expr func, expr* args, keyword*\\ & keywords, expr? starargs, expr? kwargs)]\\
        3 & \textbf{ApplyRule}[expr -\textgreater Attribute(expr value, identifier attr)]\\
        4 & \textbf{ApplyRule}[expr -\textgreater Name(identifier id)]\\
        5 & \textbf{GenToken}[my\_list]\\
        6 & \textbf{GenToken}[sort]\\
        7 & \textbf{Reduce}\\
        8 & \textbf{ApplyRule}[keyword -\textgreater keyword(identifier arg,\\ & expr value)]\\
        9 & \textbf{GenToken}[reverse]\\
        10 & \textbf{ApplyRule}[expr -\textgreater Name(identifier id)]\\
        11 & \textbf{GenToken}[False]\\
        12 & \textbf{Reduce} \\
        13 & \textbf{Reduce} \\
        14 & \textbf{Reduce} \\
        \bottomrule
    \end{tabular}
  \end{minipage}
  \caption{The AST of Python statement {\fontfamily{\ttdefault}\selectfont my\_list.sort(reverse=False)} (left) and its corresponding action sequence (right).}
\end{figure*}

A problem of converting AST to action sequence is for some code lines like function declaration or {\fontfamily{\ttdefault}\selectfont for}/{\fontfamily{\ttdefault}\selectfont while} iterators, their syntax is not complete thus cannot be parsed into ASTs. In the inference stage, if the model generates the action sequence for these statements, the tranX parser cannot convert them back to source code. When we counter these situations, we manually add the action sequence of a {\fontfamily{\ttdefault}\selectfont pass} statement after the generated sequence. Then we can convert the modified sequence back to source code and remove the added {\fontfamily{\ttdefault}\selectfont pass} statement to get the final output code.

\section{Experiments}
\subsection{Dataset}
We evaluate our approach on two public Python datasets crawled from Github repositories. The first one, Py150 \cite{raychev2016probabilistic} contains 150,000 Python2 files, split in to a training set of 100,000 files and test set of 50,000 files. In our experiments, we take 10\% of the training files as the validation set. The second dataset PyCodeSuggest \cite{bhoopchand2016learning} consists of 949 projects compatible with Python3. We create our experiment data by removing python files with action sequences longer than the maximum input length 1500 for both datasets. We list the detailed information of tokens and actions of two datasets in Table 2. We can see that program files in Py150 contain more lines, while the code lines in PyCodeSuggest are usually longer than those in Py150. For both datasets, converting source code tokens to action sequences increases the average length of code lines.

\begin{table}[htbp]
\caption{Details of the dataset.}
\begin{center}
    \begin{tabular}{lll}
    \toprule
    \multicolumn{1}{l}{} & \multicolumn{1}{l}{Py150} & \multicolumn{1}{l}{PyCodeSuggest} \\
    \midrule
    Avg. lines of code & 28.2 & 11.4\\
    Avg. tokens per statement & 10.6 & 23.6\\
    Avg actions per statement & 14.7 & 35.6\\
    \bottomrule
    \end{tabular}%
\label{tab1}
\end{center}
\end{table}

\subsection{Metrics and Baselines}
We use the following metrics to evaluate the performance of our approaches:
\begin{itemize}
    \item Accuracy: We compare the exact matching accuracy between the generated statement and the ground truth. We list the accuracy of top-1 and top-5 candidates in our paper. We further calculate the accuracy when neglecting identifier names to measure the ability to generate correct statement structures for our models.
    \item Mean reciprocal rank (MRR): we calculate MRR from the top-5 candidates of every test sample. We use it to measure the ability to give the correct sequence a high rank, which is similar to the real scenario of code completion, where developers are given a group of suggestions.
    \item BLEU \cite{papineni2002bleu}: BLEU measures the precision of N-grams, so we use it to measure the similarity between the target statement and the generated statement.
    \item Edit similarity: if the suggested code line is not precisely correct, developers want to make as few edits as possible to correct the code line. The character-level edit similarity of the predicted output $\hat{y}$ and the target output $y$ is computed by:
    \begin{equation}
        sim=1-\frac{lev(\hat{y},y)}{|y|+|\hat{y}|}
    \end{equation}
    Here $lev()$ is the Levenshtein distance (edit distance) and $\left |\cdot \right |$ is the length (number of characters) of the sequence.
\end{itemize}

We evaluate the following models on our datasets:
\begin{itemize}
    \item BPE-NLM \cite{karampatsis2020big}: a GRU language model for code completion using BPE to split source code tokens. This approach achieved state-of-the-art results on code completion tasks in multiple programming languages.
    \item Language models: we use Transformer and GRU language models on source code token sequences or ASDL action sequences. In order to analyze the effectiveness of byte-pair encoding, we also run Transformer language models on BPE subtokens.
\end{itemize}
We have also attempted to apply encoder-decoder models for this task, i.e., using an encoder to encode the program context and a decoder to generate the target statement. However, for encoder-decoder models, we have to create separate training samples for each statement in the code files, which tremendously increase the size of the training set, making the training time unacceptable.

There also exists a group of researches on code completion by predicting the next AST node in the flattened sequence of ASTs \cite{raychev2016probabilistic,li2018code,liu2019self}. Our goal is to recommend a line of code to developers, while a sequence of AST nodes cannot always be transferred into source code. So these approaches are not suitable for full-line code completion, and we do not evaluate them in our experiments.

\subsection{Experimental Setup}
In each test sample, the input is the first $k$ lines of a Python file, and the target output is the $k+1$th line. So for a Python file of $N$ lines, we can create $N-1$ test samples. We remove test samples whose target is an {\fontfamily{\ttdefault}\selectfont Import} statement or longer than 100 tokens. We exclude {\fontfamily{\ttdefault}\selectfont Import} because they are often irrelevant to the previous code context. For code statements longer than 100 tokens, we find out these statements are often the definition of large lists or dictionaries, which are incapable for neural language models to complete.

For token-based and syntax-based approaches, we set the vocabulary size to 80,000. For BPE-based approaches, we set the number of merge operations to 30,000. The hidden size of GRU models is 512. The batch size is 8 for GRU and 4 for Transformer models, since larger batches cannot be fed once into the memory. We use the Adam optimizer \cite{DBLP:journals/corr/KingmaB14} to train all models. We implement our models in PyTorch \cite{paszke2019pytorch} and run our experiments on a NVIDIA Tesla V100 GPUs with 16GB memory.

\subsection{Results and Analysis}
\begin{table*}[htbp]
  \centering
  \caption{Experiment results for all models on Py150.}
    \begin{tabular}{lrrrrrrr}
          \multicolumn{1}{l}{Model} & \multicolumn{1}{l}{acc@1} & \multicolumn{1}{l}{acc@1 w/o id} & \multicolumn{1}{l}{acc@5} & \multicolumn{1}{l}{acc@5 w/o id} & \multicolumn{1}{l}{MRR} & \multicolumn{1}{l}{BLEU-4} & \multicolumn{1}{l}{Edit similarity} \\
          \midrule
    BPE-NLM & 6.94     & 11.06     & 11.62     & 20.29     & 8.73     & 11.95     & 48.31    \\
    GRU+Token & 8.38     & 10.84     & 13.94     & 20.03     & 10.43     & 16.59     & 51.81     \\
    GRU+Syntax & 6.93     & 8.86     & 10.47     & 15.78     & 8.73     & 15.14     & 47.76     \\
    TransformerLM+Token & \textbf{8.93}     & 12.47     & \textbf{15.98}     & 23.76     & \textbf{11.47}     & \textbf{19.58}     & \textbf{54.03}     \\
    TransformerLM+Syntax & 7.73     & 10.98     & 12.68     & 20.25     & 9.97     & 19.27     & 50.90    \\
    TransformerLM+BPE & 8.55     & \textbf{14.12}     & 14.72     & \textbf{25.82}     & 10.82     & 16.11     & 52.22    \\
    \end{tabular}%
  \label{tab:addlabel}%
\end{table*}%

\begin{table*}[htbp]
  \centering
  \caption{Experiment results for all models on PyCodeSuggest.}
    \begin{tabular}{lrrrrrrr}
          \multicolumn{1}{l}{Model} & \multicolumn{1}{l}{acc@1} & \multicolumn{1}{l}{acc@1 w/o id} & \multicolumn{1}{l}{acc@5} & \multicolumn{1}{l}{acc@5 w/o id} & \multicolumn{1}{l}{MRR} & \multicolumn{1}{l}{BLEU-4} & \multicolumn{1}{l}{Edit similarity} \\
          \midrule
    BPE-NLM & 2.98     & 5.71     & 5.10     & 10.61     & 3.77     & 5.77     & 40.11    \\
    GRU+Token & 2.78     & 4.84     & 5.10     & 9.08     & 3.60     & 6.22     & 40.40     \\
    GRU+Syntax & 1.41     & 4.39     & 2.28     & 8.43     & 1.98     & 6.55     & 39.17     \\
    TransformerLM+Token & \textbf{4.32}     & 8.35    & \textbf{8.65}     & 17.05     & \textbf{5.81}     & 12.61     & \textbf{48.59}     \\
    TransformerLM+Syntax & 3.58     & \textbf{8.72}     & 6.82     & \textbf{17.64}     & 4.87     &  \textbf{14.92}    & 47.67     \\
    TransformerLM+BPE & 3.90     & 7.88     & 7.81     & 16.55     & 5.27     & 11.31     & 46.68    \\
    \end{tabular}%
  \label{tab:addlabel}%
\end{table*}%

Table 3 and Table 4 show the results of all models on the two datasets. Results on all metrics are reported in percentage (\%). We can see that in all experiment settings, Transformer models outperform GRU models on all evaluation metrics. A somewhat unexpected finding is that syntax-based approaches do not outperform token sequence-based ones, and even perform worse on Py150. We assume this is caused by the differences in sequence length between these two types of approaches. From Table 2, we can clearly figure out that the action sequence for a Python program is often longer than its source code token sequence, which brings an extra burden to language models. The main advantage of ASDL action sequences over source code tokens is that an action sequence can be necessarily converted to a syntactically correct code snippet. This is important for NL-based code generation tasks like Django \cite{oda2015learning} or CoNaLa \cite{yin2018learning} since in these tasks, the source code corpus is small-scaled and each code sample is short. However, in large-scale real-world corpora with file-level code snippets, neural language models are capable of learning program grammar from data, so the effect of explicit syntax restrictions becomes undermined. We manually inspected the statements generated by TransformerLM+token, and nearly all of them are syntactically correct. Also, we must notice that currently, our syntax guidance is still very coarse for code completion. First, we do not augment our language models with syntactical dependencies like AST parent-child connections. The neural language models are still purely sequential. Second, the ASDL parser in our experiments only guarantee that generated code lines can be parsed into ASTs, while in the real scenario of code completion, the restriction of outputs is much stronger, including restriction on variable usage, API calls, and arguments, etc. A promising way of applying these restrictions is to leverage powerful static analysis tools. Static analysis can be applied to source code tokens, which relieves language models from learning on longer ASDL action sequences.

Another phenomenon is that language models with BPE outperform token-level language models on accuracy without identifiers but achieved similar or lower results than token-level models on accuracy with identifiers. This implies that BPE can mitigate the out-of-vocabulary problem to a certain extent, but for some identifiers, token-based models can generate them with only one step, while in BPE, they are separated to subtokens, which increase the difficulty of generating them.

\begin{table}[htbp]
\caption{The inference time of all models on Py150.}
\begin{center}
    \begin{tabular}{lc}
    \toprule
    \multicolumn{1}{l}{Model} & \multicolumn{1}{c}{Avg. inference time (s)}\\
    \midrule
    BPE-NLM & 0.07\\
    GRU+Token & 0.08\\
    GRU+Syntax & 1.41\\
    TransformerLM+Token & 0.07\\
    TransformerLM+Syntax & 1.34\\
    TransformerLM+BPE & 0.08\\
    \bottomrule
    \end{tabular}%
\label{tab1}
\end{center}
\end{table}

We also make an evaluation of the time efficiency of our models. Table 5 compare the average time consuming for inferencing a whole line of code for all models. We report the time cost when the beam size is set to 5. Although all models have acceptable time efficiency, the inference speed of syntax-based models are much slower than token-based or BPE subtoken-based models. This shows that masking out invalid actions with an ASDL parser is much more time consuming than predicting a code element with neural language models. From our experiments, we believe that in order to keep a high time efficiency for full-line code completion, our model should be based on source code tokens. We need to explore the possibility of adding grammatical constraints to source token sequences.

\section{Conclusion and Future Work}
In this paper, we define a new task, full-line code completion, and studied the performance of neural language models on this task. Apart from token-based and BPE-based approaches, which have already been evaluated on token-level code completion tasks, we additional conduct experiments with ASDL syntax-based models. Our experiments show that Transformer language model on token sequences currently performs best on our datasets.

In the future, we plan to further improve the effectiveness of language models on full-line code completion by training on more data and using models with larger parameter size. Meanwhile, we aim to utilize more powerful software analyzing tools to further narrow down the output space of our model, e.g., adding restrictions on variable names and API usage. Furthermore, we would like to improve our neural model to incorporate syntax structures like parent-child links in ASTs and incorporate BPE or copy mechanism to tackle the out-of-vocabulary problem.

\bibliography{ref2}
\end{document}